\documentclass[prb,aps,twocolumn,floats,showpacs,amsmath,amssymb]{revtex4}
\usepackage[english]{babel}
\usepackage{graphicx}
\usepackage{amssymb}
\usepackage{amsmath}

\begin{document}

\title{Detecting the Berry curvature in photonic graphene}

\author{R. L. Heinisch}
\author{H. Fehske}
\affiliation{Institut f{\"ur} Physik,
             Ernst-Moritz-Arndt-Universit{\"a}t Greifswald,
             17487 Greifswald,
             Germany}

\begin{abstract}
We describe a method for measuring the Berry curvature  from the wave-packet dynamics in perturbed arrays of evanescently coupled optical waveguides with honeycomb lattice structure. 
To disentangle  the effects of the Berry curvature and the energy dispersion we utilize a difference measurement by propagating the wave packet under the influence of a constant external force back and forth. In this way a non-vanishing Berry curvature is obtained for photonic graphene with small sublattice bias or strain, where the relative error between the exact Berry curvature and  the one derived from the semiclassical dynamics is  negligible. For the strained lattice we demonstrate the robustness of the Berry curvature texture over the Brillouin zone compared to the energy dispersion.   We also comment on the experimental realization of the proposed Berry curvature mapping in photonics.
\end{abstract}

\maketitle
\section{Introduction}
The analysis of wave-packet dynamics in photonic lattices arranged in a honeycomb structure may develop and broaden the theoretical understanding of topologically protected states in both optical and quantum matter systems.  A photonic lattice is a periodic arrangement of waveguides that are coupled evanescently to each other.  Owing to optical--quantum analogies, the paraxial propagation of light through such structures can be described by a two-dimensional Schr\"odinger-type equation~\cite{RZT13,SN10}, in that
\begin{align}
i\frac{\lambda}{2\pi}\frac{\partial}{\partial z} E(x,y,z)=-\Big[\frac{\lambda^2}{(2\pi)^2}&\frac{1}{2 n_0}\Big(\frac{\partial^2}{\partial x^2}+\frac{\partial^2}{\partial y^2}\Big)\nonumber\\&+\Delta n(x,y,z\Big] E(x,y,z)
\label{phe}
\end{align}
is the optical paraxial Helmholtz equation for the evolution of the electrical field envelope  $E$.  From Eq.~(\ref{phe}), the correspondence between distance $z$ and time $t$, bulk refraction index $n_0$ and mass $m$, refractive index change $\Delta n=n_0-n(x,y,z)$ and  potential $V$, and Planck's constant $h$ and wavelength $\lambda$ is evident.  
The main advantage is that in optics the evolutions takes place in space instead of time, and therefore can be directly observed at an arbitrary resolution~\cite{SN10}. Moreover, compared to carbon-based graphene, in photonic graphene one can easily manipulate the lattice geometry and measure the dynamics of the wave function. Additional effects, e.g., due to the presence of nonlinearity, can be achieved by using Kerr media. Thus optical waveguides provide valuable, tunable model systems to emulate and study condensed matter phenomena. Indeed,  photonic lattices have been used to experimentally demonstrate a topological transition of classical light by the creation and destruction of an edge state~\cite{RZT13}, the adiabatic dynamics of edge waves~\cite{ACM15}, Landau quantisation due to strain-induced pseudo-magnetic fields~\cite{RPZ13}, topologically protected  transport of  light on the edges of Floquet topological insulators~\cite{RZP13}, anomalous and integer quantum Hall effects in driven dissipative situations influenced by a synthetic gauge field~\cite{OC14}.

In this paper  we address  the topological properties of a system of coupled optical waveguides with honeycomb geometry, and discuss the measurement of the Berry curvature~\cite{Be84}.  The Berry curvature has many important consequences in condensed matter and cold atom systems~\cite{XCN10,CC13,KSPM11,TGUJE12,JMDLUGE14,ALSABNCBG15}. In particular the geometric structure of an single Bloch  band is uniquely characterized by the distribution of the Berry curvature over the Brillouin zone~\cite{DLRBSS15,FRTVLSW15}. Our approach is based on the semiclassical wave-packet dynamics and should work for any tight-binding Hamiltonian that can be realized in an optical system. This concept is well-established~\cite{CN96,SN99,DSND03} and, for example, has been used  to analyze the Berry curvature of a honeycomb lattice of ultracold atoms~\cite{PC12,PC13}. For pristine graphene, the Berry curvature vanishes over the whole Brillouin zone as a consequence of time reversal and inversion symmetry. We  therefore consider two particular cases, graphene with a small sublattice bias and strained graphene, which both allow for a non-vanishing Berry curvature. 

\section{Semiclassical wave packet dynamics}
The semiclassical equations of motion for a wave packet under the influence of an external force is given by~\cite{CN96,SN99b}
\begin{equation}
\dot{\bf r}=\frac{\partial \varepsilon ({\bf k})}{\partial {\bf k}} - \dot{\bf k}\times {\bf \Omega}({\bf k})\,, \quad \dot{\bf k}={\bf F}\;.
\label{eq1}
\end{equation}
Obviously the semiclassical dynamics of a wave packet is controlled by the energy dispersion $\varepsilon ({\bf k})$ as well as by the Berry curvature  ${\bf \Omega}$. The effect of the dispersion depends only on the wave packet's ${\bf k}$ vector and not on the force ${\bf F}$. The Berry curvature is defined at every point in ${\bf k}$-space. It affects the dynamics of a wave packet only under the influence of a force, that is only for varying ${\bf k}$. As a consequence,  any measurement based on the wave-packet dynamics will measure the combined effect of the Berry curvature along a path in ${\bf k}$-space. If, however, this path is limited to a sufficiently small region in ${\bf k}$-space and the Berry curvature does not vary strongly over this region, we can obtain the Berry curvature directly by monitoring the movement of the wave packet.

This raises the question which path in ${\bf k}$ space (or which applied force) is best suited for revealing the Berry curvature. While this question cannot be answered in generality, let us consider---for the purpose of illustration---two particular cases. For a circular path in ${\bf k}$ space, the integral over the Berry velocity $\dot{\bf k}\times{\bf \Omega}$ adds up to zero. Consequently, a circular path is not suited for testing the Berry curvature. The opposite extreme is a small linear segment in ${\bf k}$ space. It turns out that this path can be used for our measurement proposal. 
 
In order to disentangle the effects of the energy dispersion and the Berry curvature, we suggest a difference measurement based on the propagation of a wave packet under a constant force compared to the propagation under the reversed force.  That such kind of time-reversal protocal can been used to map out the Berry curvature has been demonstrated in the context of ultracold atom gases~\cite{PC12}. For a constant force, ${\bf F}=F{\bf \hat{e}}_\parallel$, the equations of motion read $\dot{\bf k}=F{\bf \hat{e}}_\parallel$ and $\dot{\bf r}=\frac{\partial \varepsilon({\bf k})}{\partial {\bf k}}+ F {\Omega}({\bf k}) {\bf \hat{e}}_\perp $. Assuming that both the gradient of the energy dispersion and the Berry curvature vary negligibly over the integration path, we obtain for the position of the wave packet under the external force (index $+$) or the reverse force (index $-$): 
${\bf r}^\pm(t)={\bf r}(0)+\left[\frac{\partial \varepsilon}{\partial {\bf k}}  \pm F {\Omega}({\bf k}) {\bf \hat{e}}_\perp\right] t$.
From this, we can derive the Berry curvature and the gradient of the energy dispersion as follows:  
\begin{equation}
{ \Omega}({\bf k})=\frac{({\bf r}^+-{\bf r}^-){\bf \hat{e}}_\perp}{2Ft}\,, \quad \frac{\partial \varepsilon}{\partial {\bf k}}=\frac{{\bf r}^++{\bf r}^--2{\bf r}(0)}{2t}\,.
\label{eq2}
\end{equation}

\section{Berry curvature for  photonic graphene}
As a precondition for any protocol  that attempts the measurement the Berry phase in photonic graphene the inversion symmetry of the honeycomb lattice has to be broken so that the Berry curvature is not restricted to  the singular $K$ and $K'$ points but spreads in the Brilloun zone. This can be achieved in the manner descibed below. 
\subsection{Biased photonic graphene }
First, let us  consider a graphene lattice with sublattice bias. The ${\bf k}$-dependent Hamiltonian for such an asymmetric honeycomb lattice is given by~\cite{FPG10}  
\begin{align}
H={
\;\Delta\quad \;f({\bf k}) \choose f^\ast({\bf k}) \;\; -\Delta 
}\,,
\end{align}
where the function 
$f({\bf k})=-e^{-i{\bf k}\boldsymbol{\delta_1}}-e^{-i{\bf k}\boldsymbol{\delta_2}}-e^{-i{\bf k}\boldsymbol{\delta_3}}$
 is obtained as a sum over hopping amplitudes in a tight-binding description with the nearest-neighbor (NN) vectors $\boldsymbol{\delta_1}={ \sqrt{3}/2\choose1/2}$, $\boldsymbol{\delta_2}={-\sqrt{3}/2\choose1/2}$, and $\boldsymbol{\delta_3}= {0\choose-1}$, in units of the NN distance. Here and in what follows all energies are measured with respect to the NN transfer amplitude. Note that the on-site energy $\Delta$ does not depend on ${\bf k}$ and breaks inversion symmetry. The energy $\varepsilon_\alpha$ is given by 
 \begin{align}
 \varepsilon_\alpha=\alpha \sqrt{|f|^2+\Delta^2}
 \end{align} with the band index $\alpha=\pm 1$, and 
 \begin{align}
 |f|^2=3+2\cos({\bf k}{\bf d_{12}})+2\cos({\bf k}{\bf d_{23}})+2\cos({\bf k}{\bf d_{31}})\,,
 \end{align} 
 where ${\bf d_{ij}}=\boldsymbol{\delta_i}-\boldsymbol{\delta_j}$.
For practicality, we rewrite the Hamiltonian as 
\begin{equation}
H=|\varepsilon | \left(\begin{matrix}
\cos \beta &  e^{-i\theta} \,\sin \beta \\ e^{i \theta}\, \sin \beta  & -\cos \beta 
\end{matrix}\right)\;,
\label{eq3}
\end{equation}
where $e^{-i\theta}=f/|f|$, $\cos \beta =\Delta/|\varepsilon |$,  $\sin \beta =|f|/|\varepsilon|$, and $|\varepsilon|^2=|f|^2+|\Delta |^2$. Then the eigenvectors are 
\begin{align}
|u_{{\bf k},1}\rangle=&\left(\cos(\beta/2), \sin(\beta/2)e^{i\theta} \right)^T\;,\\
|u_{{\bf k},-1}\rangle=&\left(-\sin(\beta/2)e^{-i\theta},\cos(\beta/2) \right)^T\;.
\end{align} 
From these we can derive the vector-valued Berry connection ${\bf A}_\alpha ({\bf k})=i\langle u_{{\bf k},\alpha}| \nabla_{\bf k} u_{{\bf k},\alpha} \rangle$, and the Berry vector potential $\Omega_{xy,\alpha}({\bf k})=\partial_{k_x}A_{y,\alpha}-\partial_{k_y}A_{x,\alpha} $~\cite{XCN10}. For the specific case of the asymmetric strained honeycomb lattice the Berry connection is 
\begin{align}
{\bf A}_\alpha({\bf k})=-\alpha \sin^2 (\beta/2)
{\partial_{k_x} \theta \choose\partial_{k_y} \theta}\,,
\end{align}
and the Berry curvature follows as 
\begin{align}
{\bf \Omega}_\alpha ({\bf k})\!=\!\alpha \frac{\sqrt{3}\Delta}{|\varepsilon|^3}\sin\left({\bf k}{\bf d_{23}}/2\right)\sin\left({\bf k}{\bf d_{31}}/2\right)\sin\left({\bf k}{\bf d_{12}}/2\right) {\bf \hat{e}}_z .
\label{eq4}
\end{align}

\begin{figure}[t]
\rotatebox{270}{\includegraphics[width=.49\linewidth]{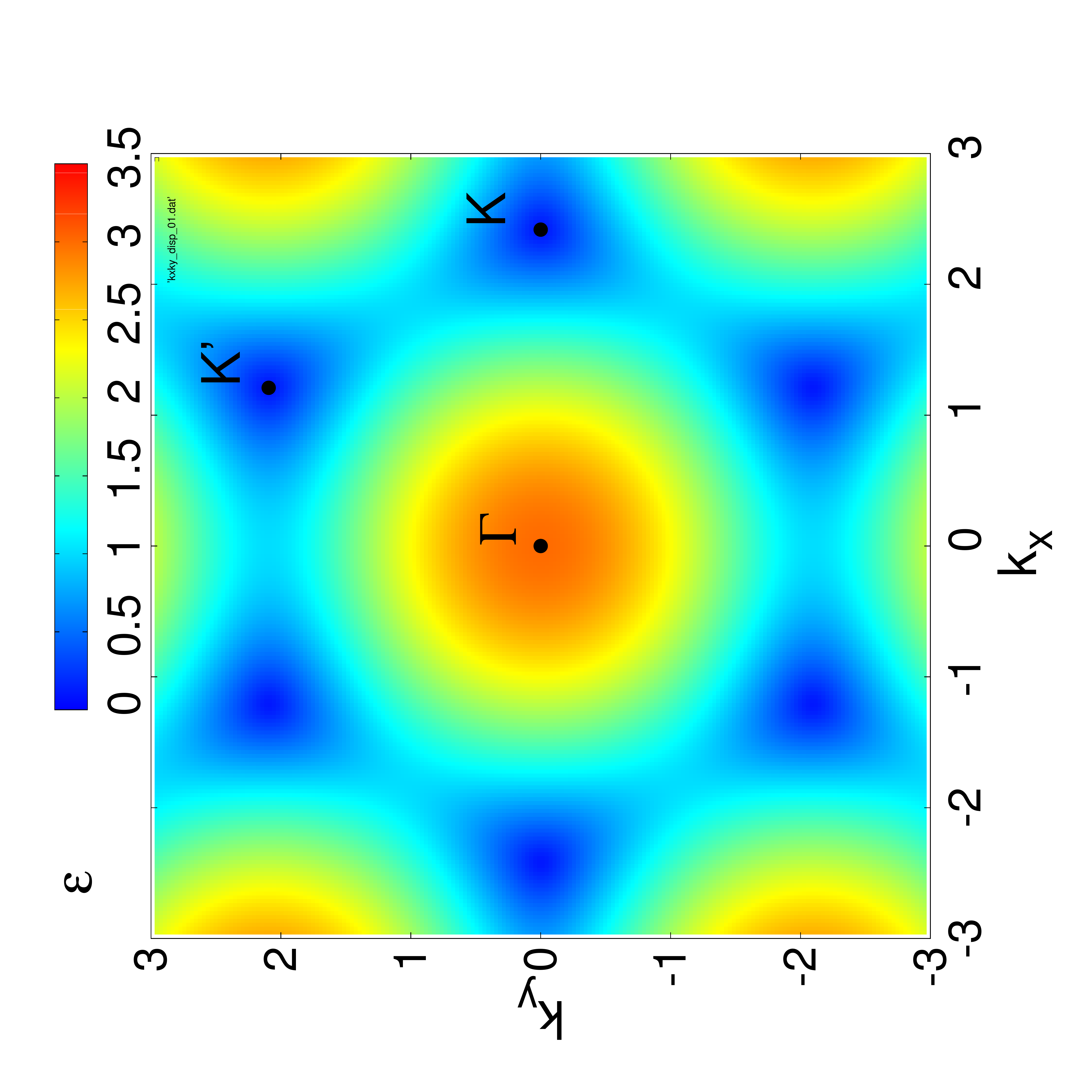}}
\rotatebox{270}{\includegraphics[width=.49\linewidth]{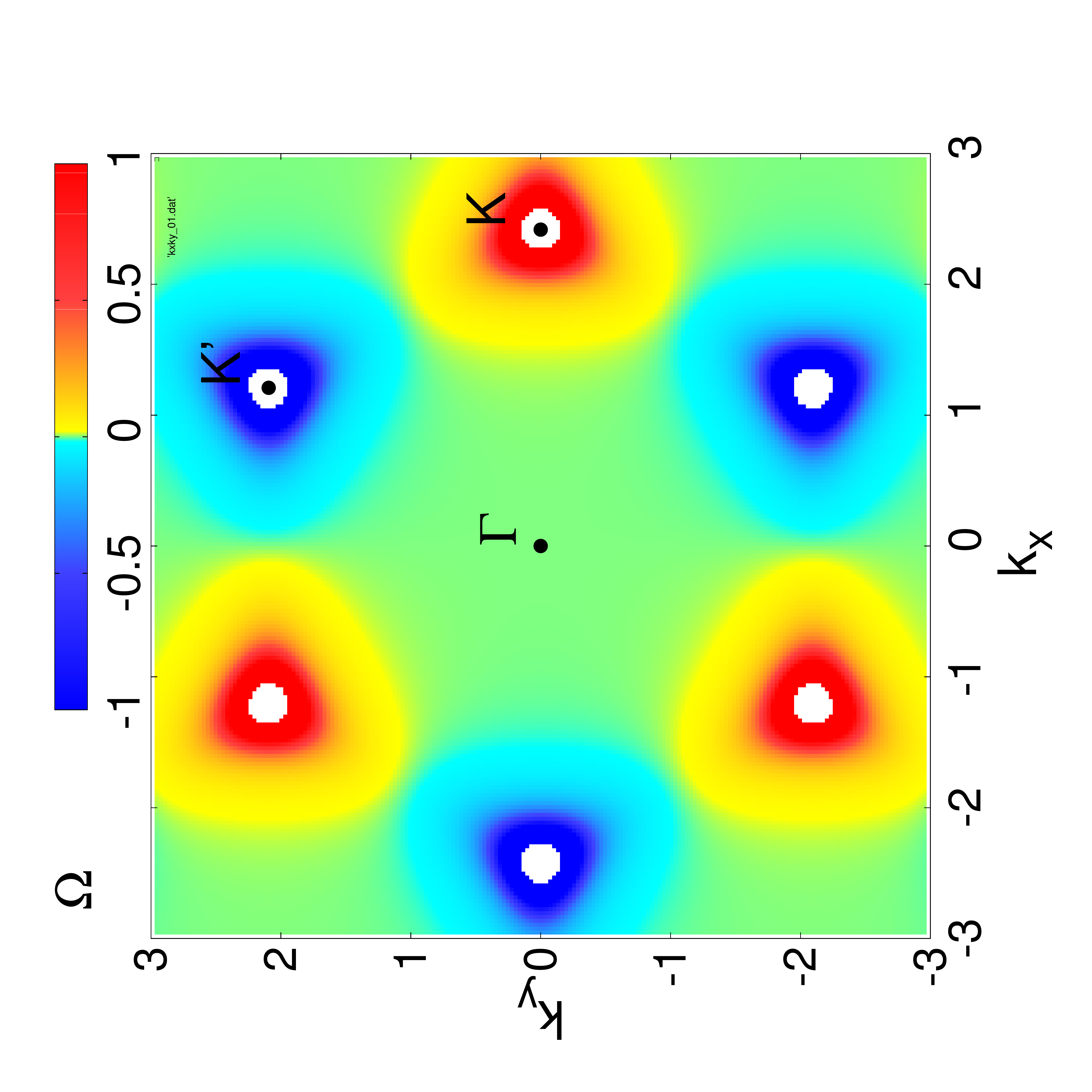}}
\caption{(color online). Energy dispersion $\varepsilon ({\bf k})$ (left panel) and Berry curvature ${\Omega} ({\bf k})$ (right panel) of the asymmetric graphene lattice for $\alpha=1$ and $\Delta=0.1$. Breaking inversion (or time-reversal) symmetry entails a non-vanishing Berry curvature throughout the Brillouin zone.}
\label{fig1}
\end{figure}

Fig.~\ref{fig1} shows the energy dispersion and the Berry curvature of photonic graphene with 
an asymmetric honeycomb lattice structure. As a consequence of the sublattice potential the system 
features two energy bands which are separated by an energy gap of of $2\Delta$ at the (nonequivalent) 
$K$ and $K'$ points located at the corners of the first Brillouin zone. At these points the Berry curvature reaches its maxima or minima~\cite{PC12}. Of course, the integral of the Berry curvature over the whole Brillouin zone vanishes, i.e., the Chern number is zero. For $\Delta \rightarrow 0$ the minima in the band structure touch the maxima of the lower energy band at zero energy, and form the Dirac cones at the K and K' points which are responsible for many of graphene's exceptional properties~\cite{CGPNG09}. In this limit the Berry curvature vanishes in the entire Brillouin zone but for the Dirac points where it becomes singular.

\begin{figure}[t]
\includegraphics[width=.9\linewidth]{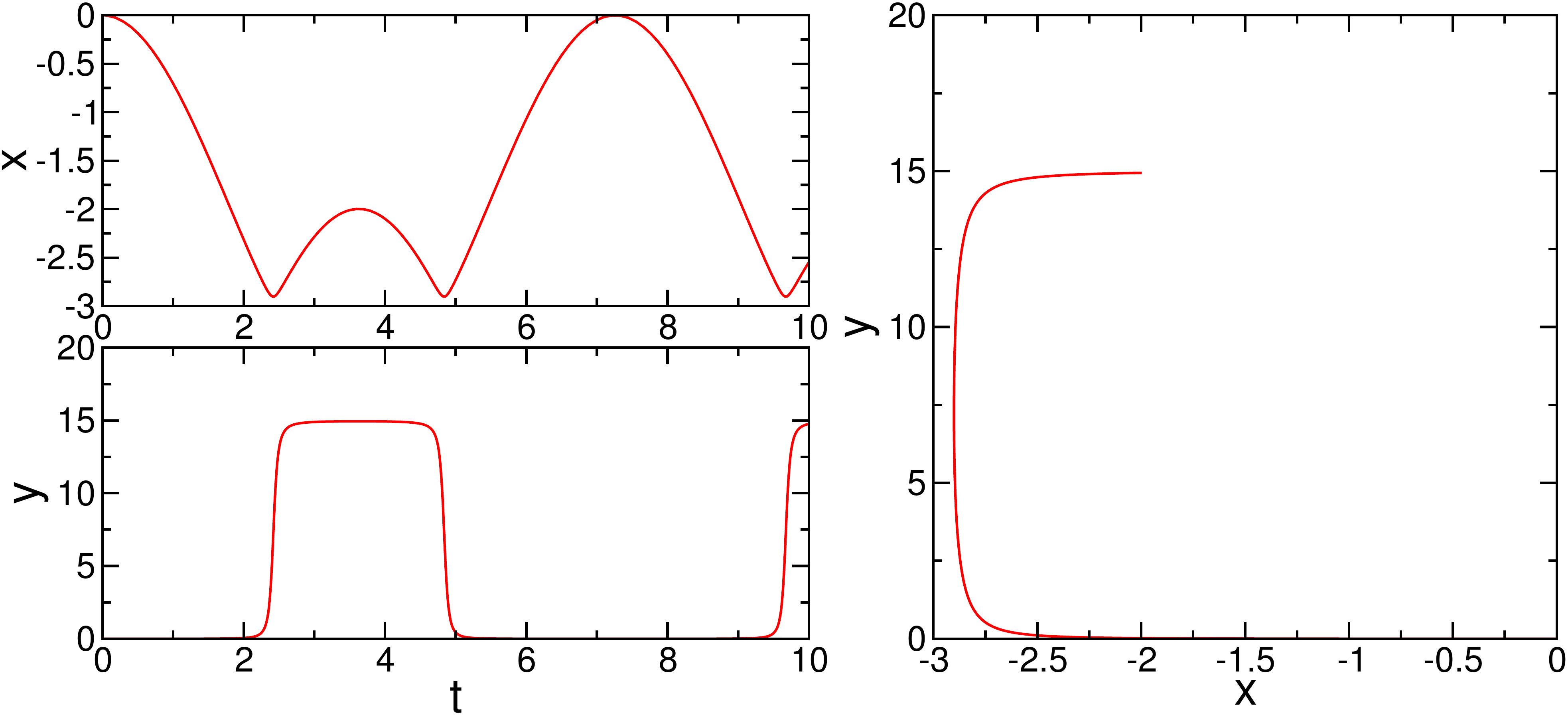}
\includegraphics[width=.9\linewidth]{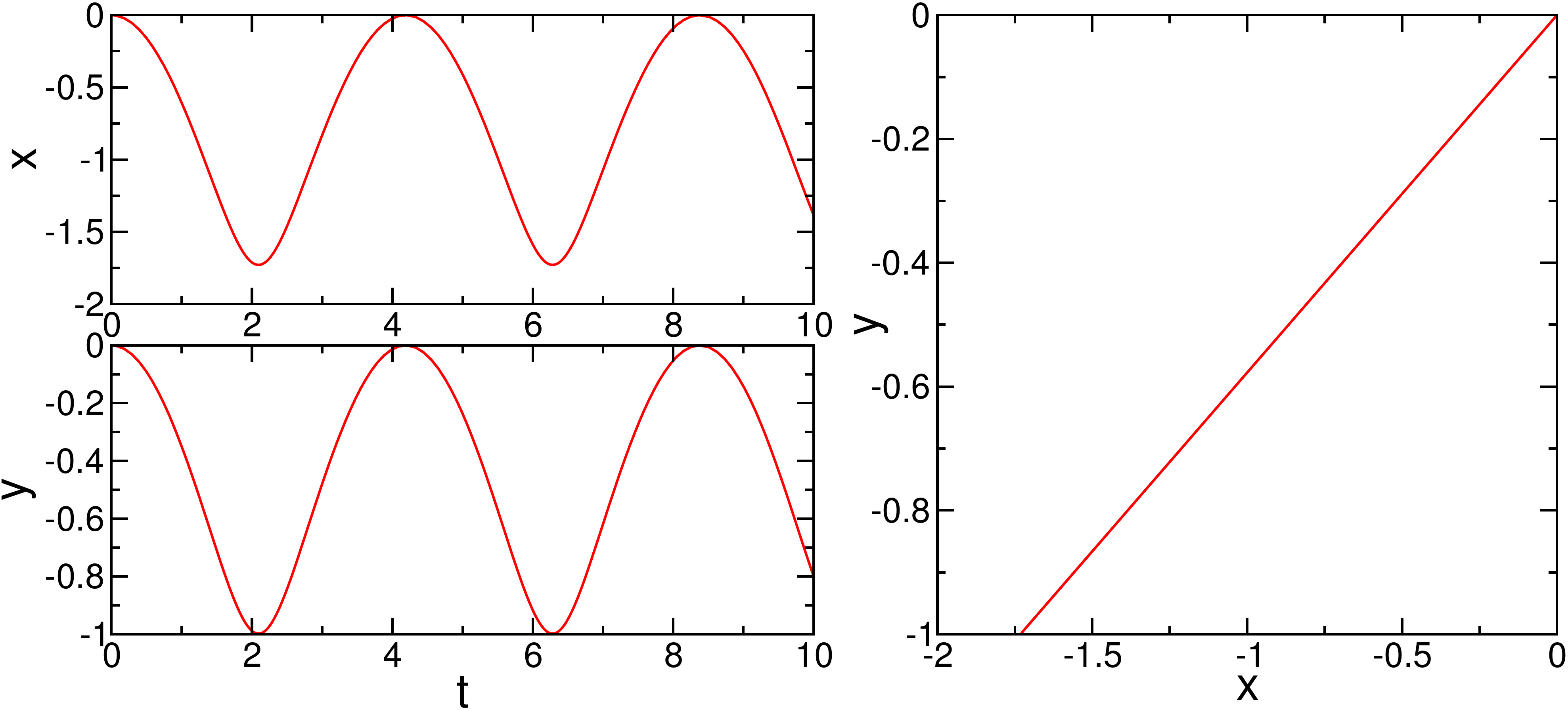}
\includegraphics[width=.9\linewidth]{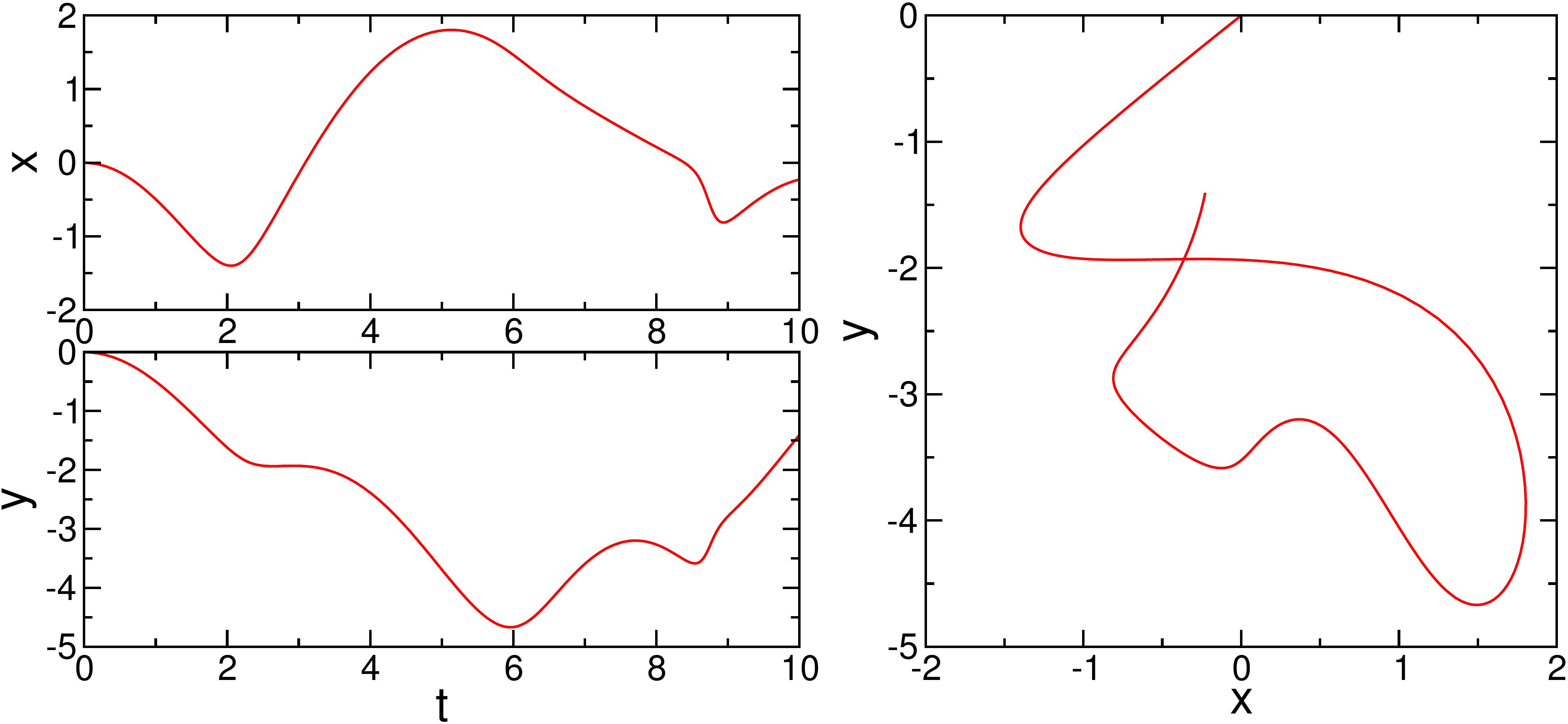}
\caption{(color online). Trajectories of a wave packet on the asymmetric graphene lattice under the action of a constant force. Each set of left panels shows the separate evolution of the $x$ and $y$ coordinates as a function of time, while the corresponding right panels give the trace of the wave packet in the $xy$ plane. The force is  $\mathbf{F}=1 \mathbf{e}_r$, where $\mathbf{e}_r(\phi)$ is the radial unit vector for $\phi=0$ (top set of panels), $\phi=\pi/6$ (middle panels), and $\phi=\pi/4$ (bottom panels). All wave packets are prepared at $x=0$, $y=0$. Again $\alpha=1$ and $\Delta=0.1$.}
\label{fig2}
\end{figure}

We now analyze the dynamics of a wave packet in the upper band, prepared at the origin in real and wave-vector space. Fig.~\ref{fig2} illustrates how its trajectory is affected by the energy dispersion and Berry curvature. The upper panel shows the effect of an external force in $x$ direction ($\phi=0$): The wave vector is pushed from the $\Gamma$ point outwards to the $K$ point,  and later through a $K'$ point back to the $\Gamma$ point. Along its path the gradient of the energy dispersion, which is always parallel to the $x$ direction, leads to a back and forth movement along ${\bf \hat{e}}_x$. The movement of the wave packet in $y$ direction is caused solely by the Berry curvature. When the wave packet passes through $K$, the positive Berry curvature leads to a rapid displacement in $y$ direction which is fully reversed once the wave packet passes through the $K'$ point where the Berry curvature is negative. Note that a similiar result was obtained for an optical lattice setup~\cite{PC12}. In the middle panel, the force points in the direction between the $K$ and $K'$ points ($\phi=\pi/6$).  Here the Berry curvature is zero. Now the back- and forth-movement of the wave packet stems only from the energy dispersion. The bottom panel shows a general situation when the influence of the energy dispersion and of the Berry curvature cannot be easily disentangled. Accordingly a rather complex trajectory of the wave packet results which encompasses the effects of both.  
  
\begin{figure}[t!]
\centering
\rotatebox{270}{\includegraphics[width=.49\linewidth]{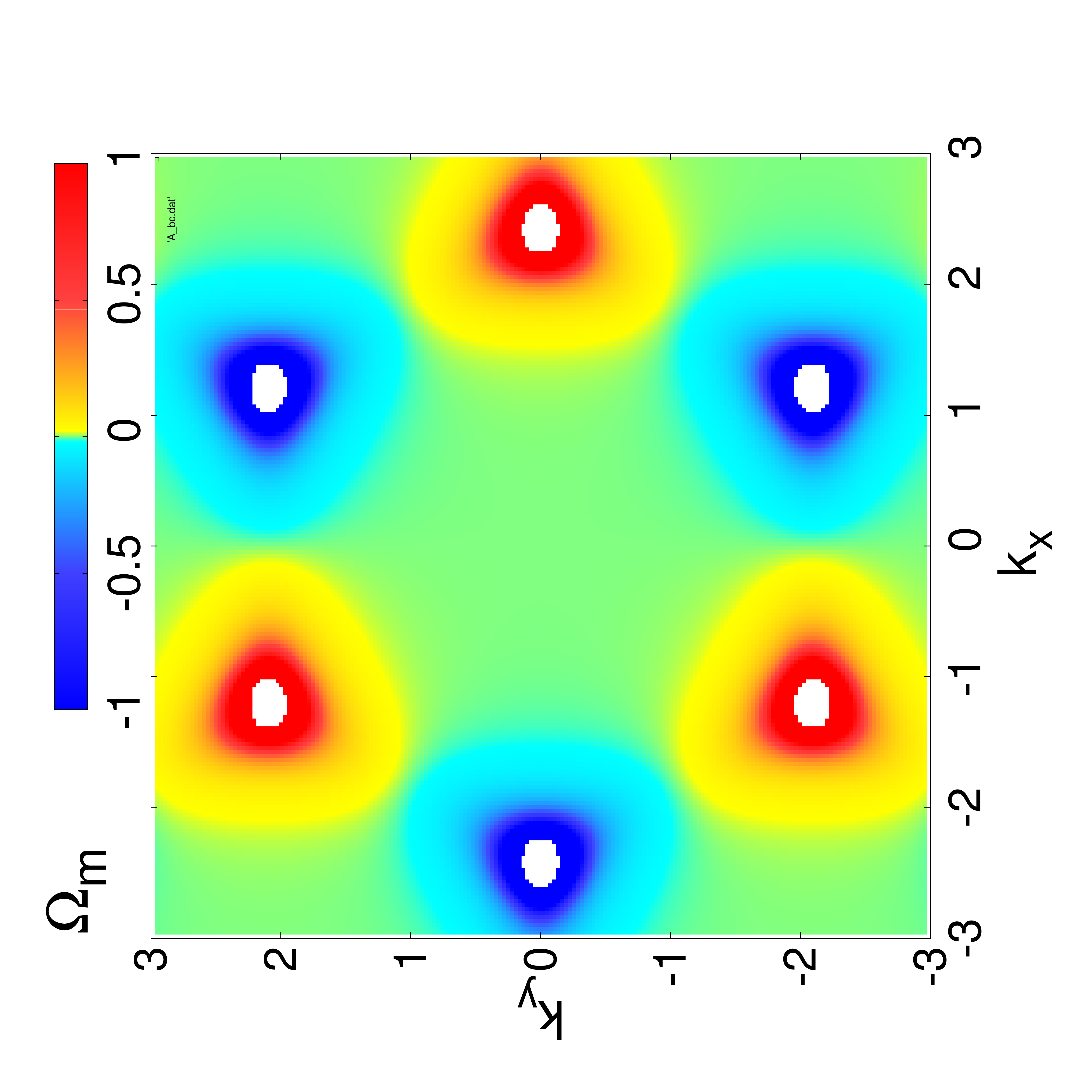}}
\rotatebox{270}{\includegraphics[width=.49\linewidth]{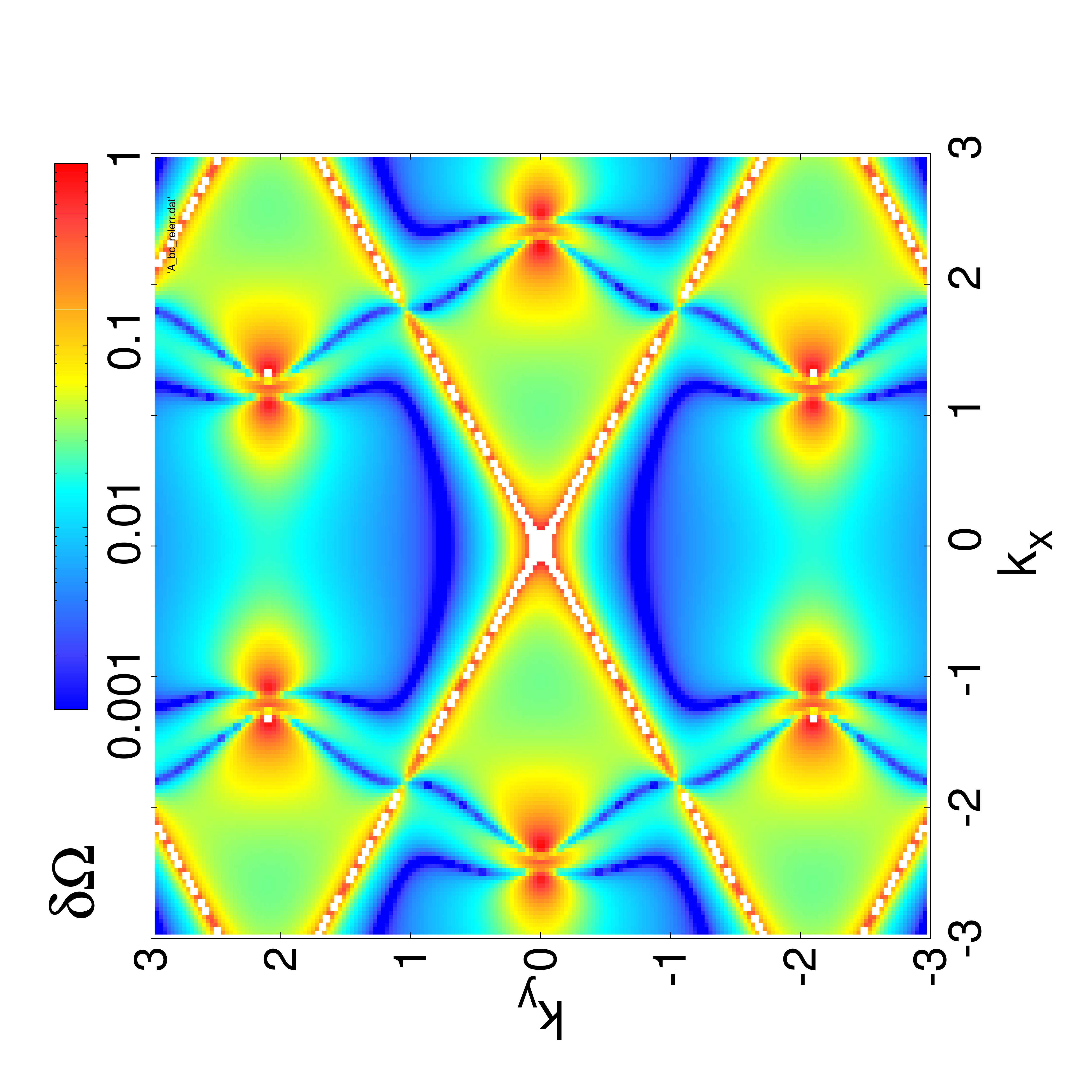}}\\
\rotatebox{0}{\includegraphics[width=.5\linewidth]{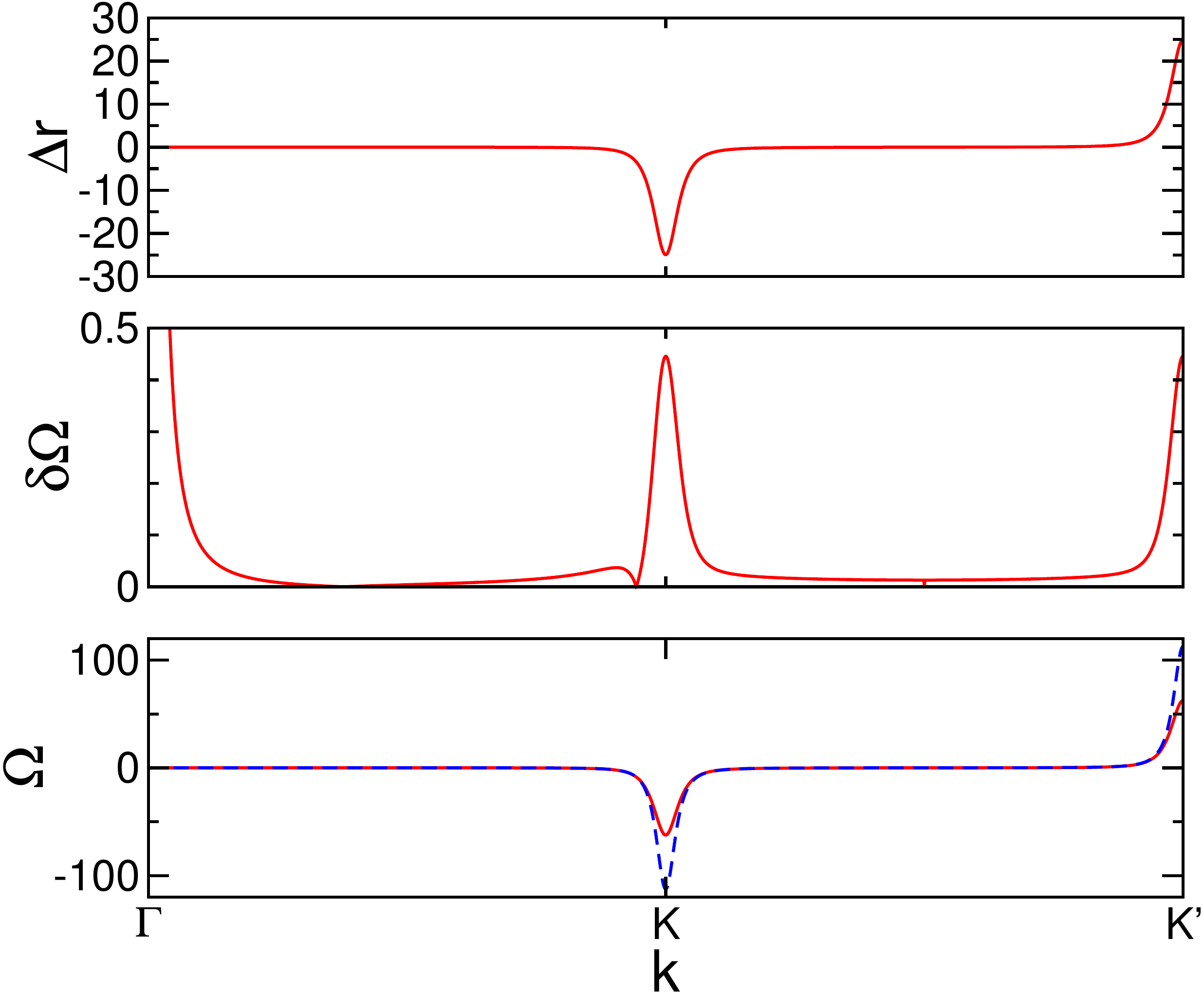}}
\caption{(color online).  Top: Mapped Berry curvature $\Omega_m$ (left) and relative deviation of $\Omega_m$ from the exact Berry curvature $\Omega_e$ (right). Bottom: Displacement of the wave packet perpendicular to the force  $\Delta r$ (upper panel), relative deviation of $\Omega_m$ from $\Omega_e$ (middle panel),
 $\Omega_m$  (red solid line) and $\Omega_e$   (blue dashed line) (lower panel), where the ordinate follows a path through the Brillouin zone from $\Gamma$  via $K$ to $K^\prime$. Here the force is $F=1{\bf \hat{e}}_x$, and the propagation time $t=0.2$.}
\label{fig3}
\end{figure}

Based on these findings we can now explore the idea of a difference measurement of the Berry curvature from wave-packet trajectories for the particular case of the asymmetric honeycomb lattice.  Apparently, the wave-packet trajectories displayed in Fig.~\ref{fig2}  reveal a strong dependence on the Berry curvature, leading to wave-packet displacements over many `lattice sites'. In Fig. \ref{fig3} we address the inverse problem: How the Berry curvature can be obtained from the wave-packet's trajectory? To this end, we consider the following situation. A wave packet, prepared with certain wave numbers $k_x$ and $k_y$, propagates under a force (parallel to ${\bf \hat{e}}_x$) for a defined period of time. Then, the same wave packet propagates under the influence of the reversed force for the same time. The upper left panel of Fig.~\ref{fig3} illustrates the final outcome, the mapped Berry curvature, of such a measurement. Clearly, the characteristic pattern of the Berry curvature in the Brillouin zone is reproduced correctly. The upper right panel gives the relative error compared to the exact Berry curvature [calculated via Eq.~\eqref{eq4}]. In line with expectation, this error is largest near the $\Gamma$ point and along lines where the Berry curvature is small. Also at the $K$ and $K'$ points the error is non-negligible. Here the Berry curvature varies rapidly, which is difficult to reproduce accurately in an integrated measurement. Note, however, that the direction of the force leads to an anisotropy of the error. For instance, to the left and right of the $K$ or $K'$ points, the force in $x$ direction causes the wave vector to sweep through a rapid variation of the Berry curvature (as well as the dispersion), entailing larger errors, whereas above or below the $K$ or $K'$ points both quantities vary only slowly in $x$ direction, yielding a relatively small error. The lower panel compares the displacement of the wave packet from which the Berry curvature is derived, the relative error and the Berry curvature along a path $\Gamma$-$K$-$K'$. Note that  the displacement of several tens of lattice sites near the interesting features of the Berry curvature at the $K$ and $K'$ points should be sufficient to be experimentally observable. 

\subsection{Strained photonic graphene}
Second, let us consider the case of a strained asymmetric honeycomb lattice, where the increased hopping amplitude in one direction compared to the other can lead to dramatic changes in the band structure. In condensed matter, the electronic spectrum of such two-dimensional crystals  
was extensively studied, e.g., with a theoretical focus on the merging of Dirac points~\cite{KSPM11,MPFG09a,MPFG09b} and a subsequent microwave tight-binding analogue experiment~\cite{BKMM13}.  

In this case we have $f(k)=-e^{-i{\bf k}\boldsymbol{\delta_1}}-e^{-i{\bf k}\boldsymbol{\delta_2}}-x'e^{-i{\bf k}\boldsymbol{\delta_3}}$, i.e., 
\begin{align}
|f|^2=&72+x'^2+2\cos({\bf k d_{12}})+2x'\cos({\bf k d_{23}})\nonumber\\&+2x'\cos({\bf k d_{31}})\,,
\end{align} 
with $x'$ being the ratios of the transfer amplitudes.  
Therewith the Berry curvature results as
\begin{align}
&{\bf \Omega}_\alpha ({\bf k})=-\frac{\alpha \Delta}{2|\epsilon|^3|f|^2} \Big\lbrace \Big[-\sqrt{3}\sin({\bf k d_{12}})+\frac{\sqrt{3}}{2}x'\sin({\bf k d_{23}})\nonumber\\&\;\;+\frac{\sqrt{3}}{2}x'\sin({\bf k d_{31}})\Big]  
\Big[(1-x'^2) +\cos({\bf k d_{12}}) -\frac{x'}{2} \cos({\bf k d_{31}})\nonumber\\
&\;\; -\frac{x'}{2} \cos({\bf k d_{23}})  \Big] -\frac{3\sqrt{3}x'^2}{4} \Big[-\sin({\bf k d_{23}})+\sin({\bf k d_{31}}) \Big] \nonumber\\&\;\;\times\Big[ \cos({\bf k d_{31}}) - \cos({\bf k d_{23}}) \Big] \Big\rbrace\,{\bf \hat{e}}_z \,.
\label{eq5}
\end{align}

\begin{figure}[h]
\centering
\rotatebox{270}{\includegraphics[width=.49\linewidth]{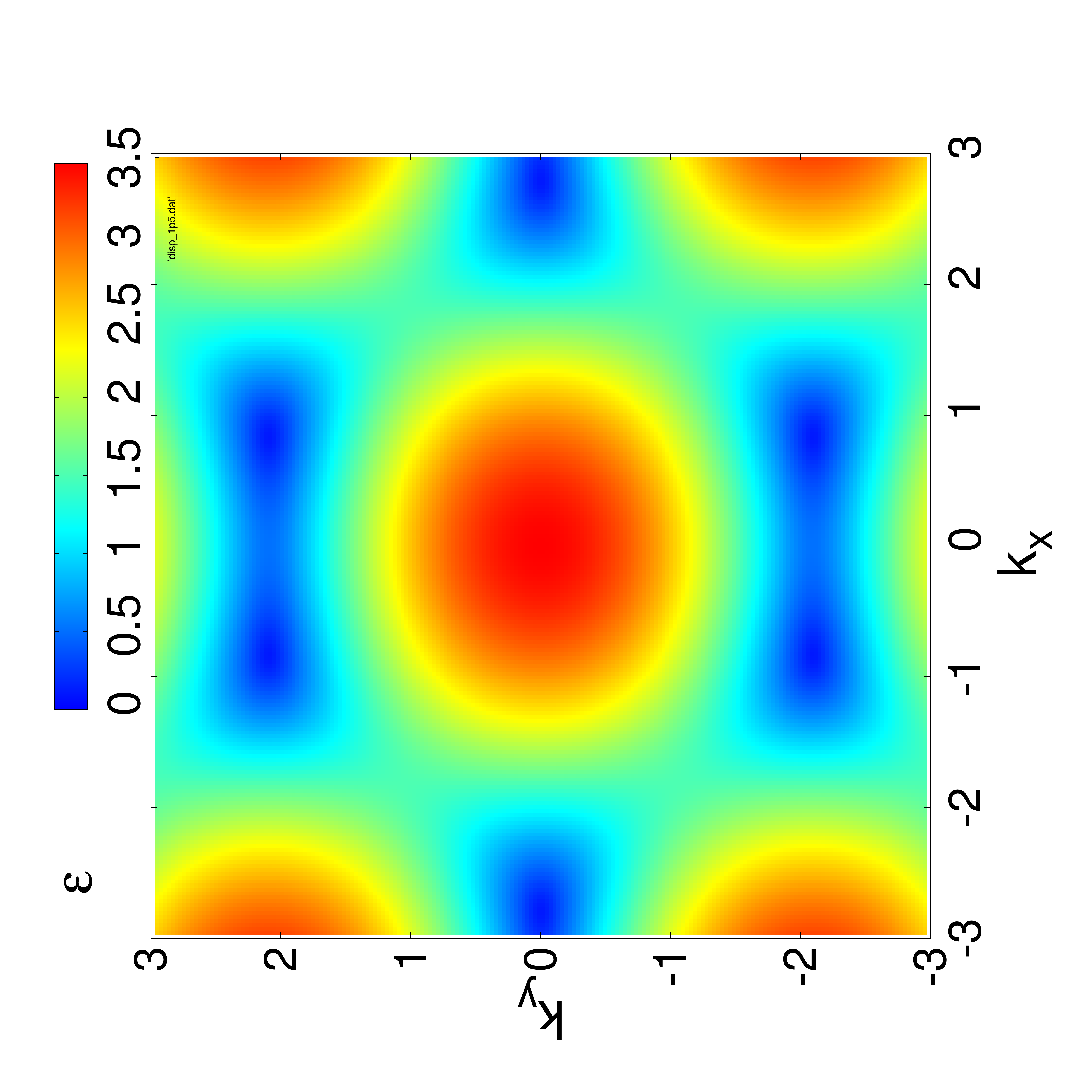}}
\rotatebox{270}{\includegraphics[width=.49\linewidth]{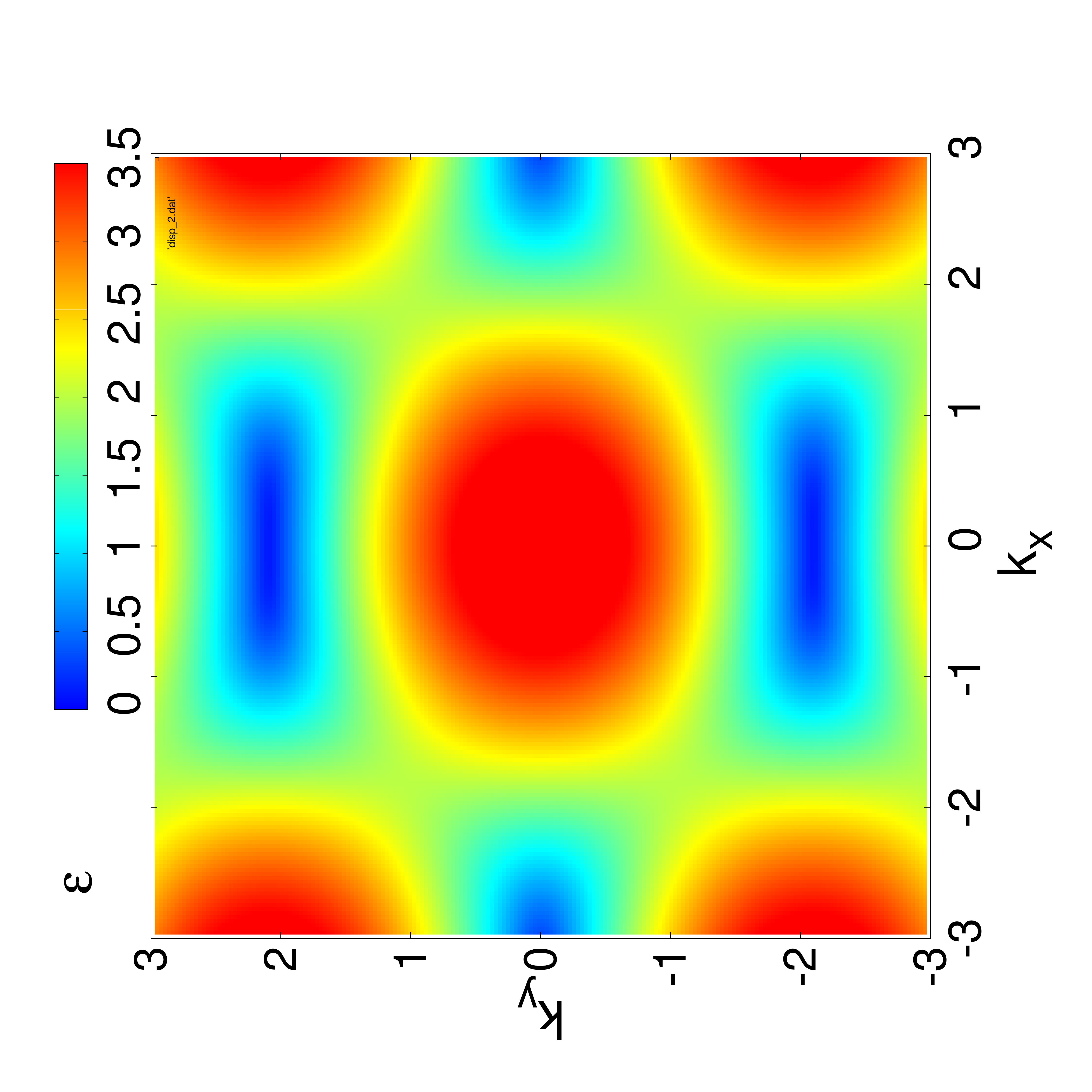}}\\
\rotatebox{0}{\includegraphics[width=.42\linewidth]{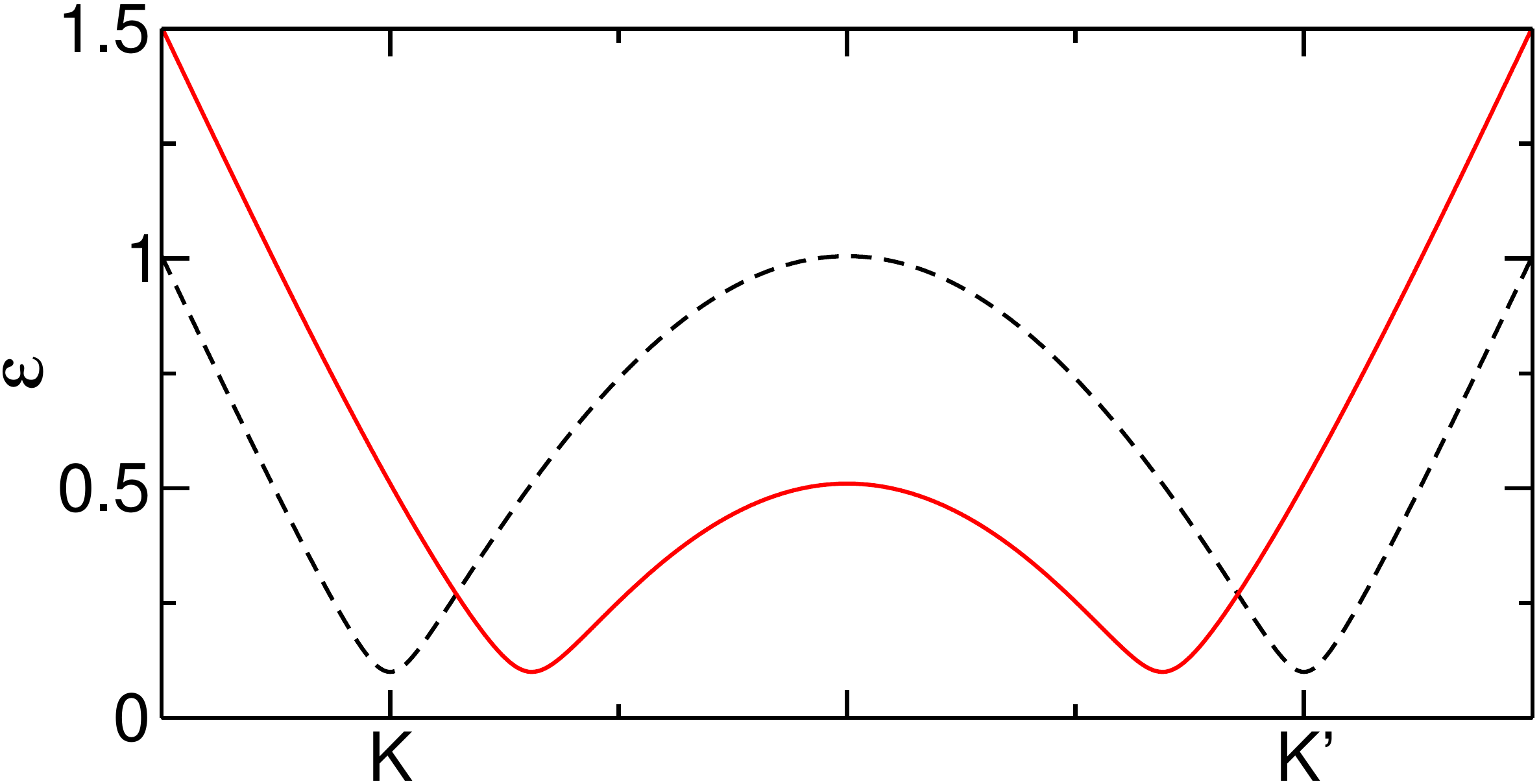}}\hspace*{0.2cm}
\rotatebox{0}{\includegraphics[width=.42\linewidth]{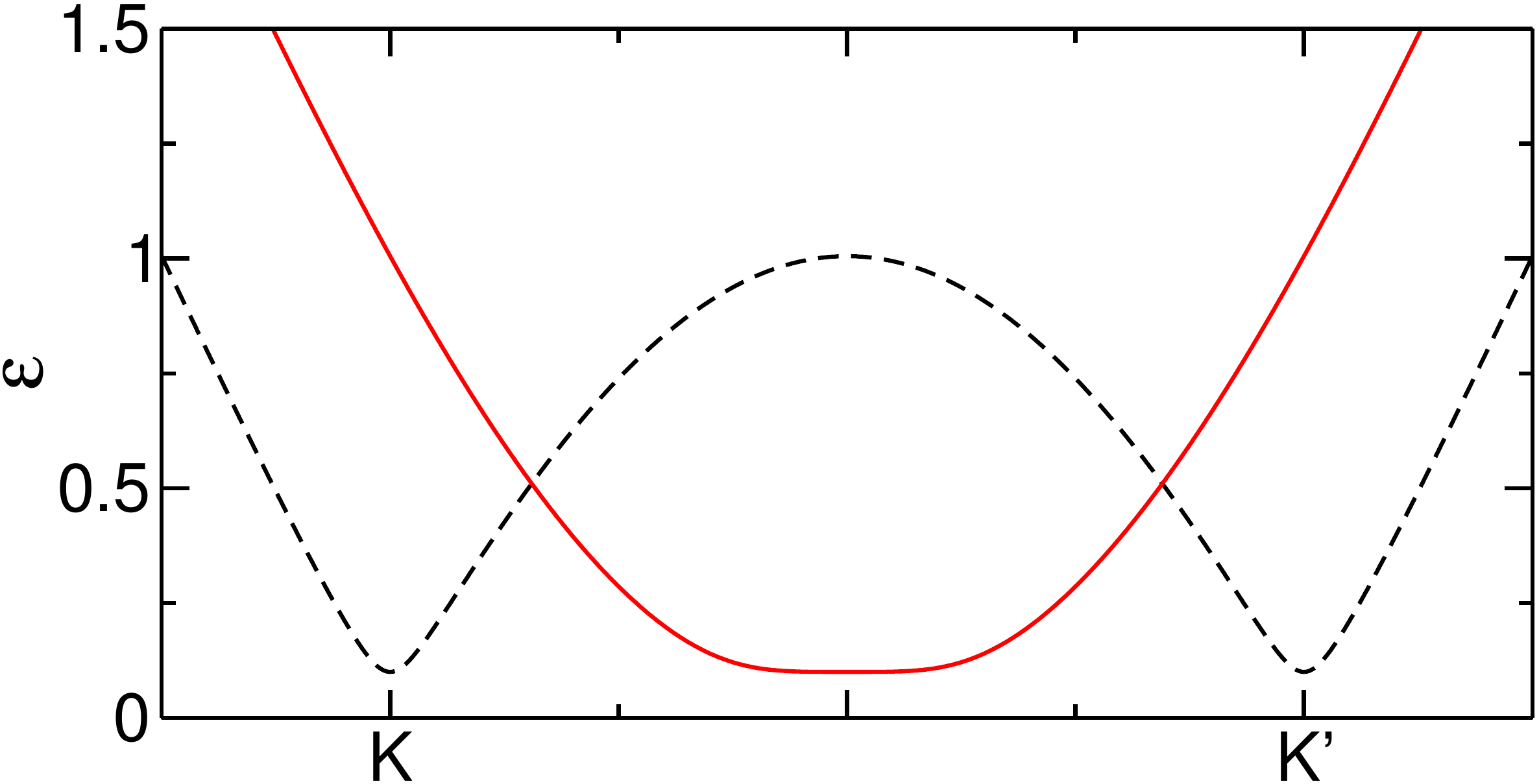}}
\hspace*{0.3cm}\\
\rotatebox{270}{\includegraphics[width=.49\linewidth]{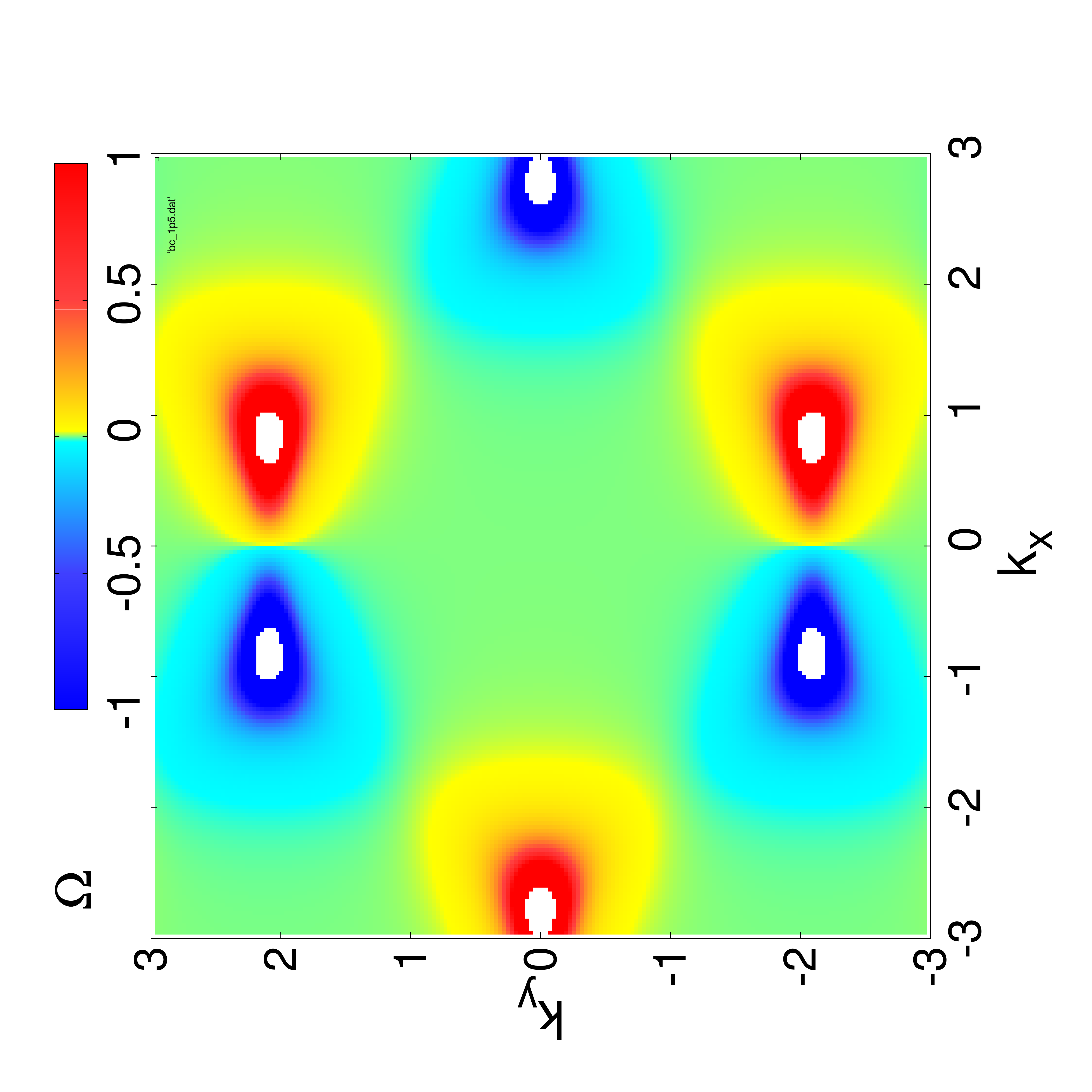}}
\rotatebox{270}{\includegraphics[width=.49\linewidth]{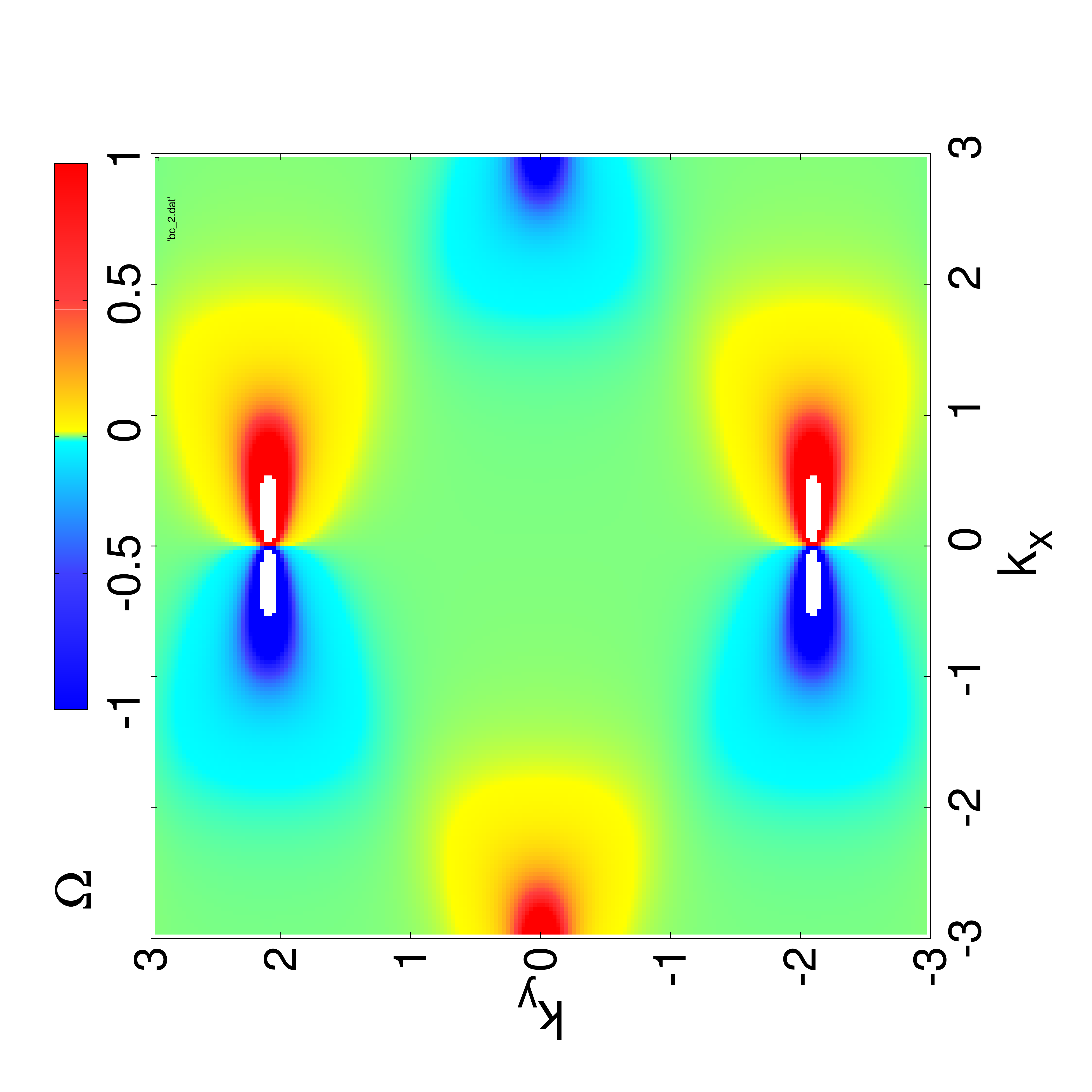}}\\
\caption{(color online). Energy dispersion 
$\varepsilon ({\bf k})$ (top) and of the Berry curvature $\Omega({\bf k})$  (bottom) for asymmetric strained photonic graphene. The middle panels give the band dispersion along a path parallel to the $k_x$ axis between neighboring $K$ and $K'$ points (red lines). The strain parameter is $x'=1.5$ (left) and $x'=2$ (right).  Contrary to the energy dispersion the characteristic features of the Berry curvature, the positive and negative peak associated with the band minima of the unstrained lattice, persist even for large strain.}
\label{fig4}
\end{figure}

Figure \ref{fig4} indicates that the energy dispersion of the strained lattice shows almost no traces of the once hexagonal symmetry. With increasing strain, adjacent band minima in $x$ direction, originally located at $K$ and $K'$  (compare the unstrained 
case $x'=1$  included in the middle panels by black dashed lines), move closer together until they merge at about  $x'= 2$. This merging was observed experimentally for a ultracold  Fermi gas in a tunable optical honeycomb lattice~\cite{TGUJE12}.   
The features of the Berry curvature, on the contrary, are relatively robust against strain. The positive and negative peaks, associated with the $K$ and $K'$ points respectively, move closer together in $x$ direction but do not merge and remain clearly distinct even for large strain. Thus the signatures of the hexagonal symmetry remain observable. The example of the strained honeycomb lattice shows how robust the Berry curvature texture over the Brillouin zone can be compared to the energy dispersion. As photonic analogues of this lattice can be manufactured, it would represent an interesting test-case for our proposal of the Berry curvature measurement. 

\section{Conclusions}
To sum up, we  theoretically demonstrated  Berry-phase effects in strained or sublattice-biased honeycomb photonic lattices. This is an optical analogue of what may be observed in carbon-based graphene. For coupled optical waveguides the
system parameters (on-site potentials, nearest-neighbor transfer amplitudes etc.) are readily modifiable, however, and (external) forces can be put into effect by manipulating the index of refraction or bending the optical fibers~\cite{SN10}. Taking the advantage of these systems, and mapping, in view of the similiarity of the Schr\"odinger equation and the paraxial Helmholtz equation, the temporal evolution onto a spatial dimension, we discuss a direct measurement of the Berry curvature based on a bidirectional wave-packet propagation in the presence of a constant force which separates the Berry curvature and band dispersion effects. Thereby the length of the waveguide has to be correlated with the propagation time. Experimentally the difference measurement might be performed by using the same waveguide from two different ends. The analysis of wave propagation in such photonic structures might become a future, versatile tool to analyze topological aspects of semiclassical and quantum matter.  

\section*{Acknowledgements}
The authors would like to thank J. M. Zeuner for valuable discussions. This work was partly funded by the Deutsche Forschungsgemeinschaft through the Priority Program SPP 1459 `Graphene'. 

%\section*{References}
%\bibliographystyle{prop2015}
\bibliography{ref}

\begin{thebibliography}{26}
\expandafter\ifx\csname natexlab\endcsname\relax\def\natexlab#1{#1}\fi
\expandafter\ifx\csname bibnamefont\endcsname\relax
  \def\bibnamefont#1{#1}\fi
\expandafter\ifx\csname bibfnamefont\endcsname\relax
  \def\bibfnamefont#1{#1}\fi
\expandafter\ifx\csname citenamefont\endcsname\relax
  \def\citenamefont#1{#1}\fi
\expandafter\ifx\csname url\endcsname\relax
  \def\url#1{\texttt{#1}}\fi
\expandafter\ifx\csname urlprefix\endcsname\relax\def\urlprefix{URL }\fi
\providecommand{\bibinfo}[2]{#2}
\providecommand{\eprint}[2][]{\url{#2}}

\bibitem[{\citenamefont{Rechtsman
  et~al.}(2013{\natexlab{a}})\citenamefont{Rechtsman, Zeuner, Tunnermann,
  Nolte, Segev, and Szameit}}]{RZT13}
\bibinfo{author}{\bibfnamefont{M.}~\bibnamefont{Rechtsman}},
  \bibinfo{author}{\bibfnamefont{J.}~\bibnamefont{Zeuner}},
  \bibinfo{author}{\bibfnamefont{A.}~\bibnamefont{Tunnermann}},
  \bibinfo{author}{\bibfnamefont{S.}~\bibnamefont{Nolte}},
  \bibinfo{author}{\bibfnamefont{M.}~\bibnamefont{Segev}}, \bibnamefont{and}
  \bibinfo{author}{\bibfnamefont{A.}~\bibnamefont{Szameit}},
  \bibinfo{journal}{Nature Phys.} \textbf{\bibinfo{volume}{7}},
  \bibinfo{pages}{153} (\bibinfo{year}{2013}{\natexlab{a}}).

\bibitem[{\citenamefont{Szameit and Nolte}(2010)}]{SN10}
\bibinfo{author}{\bibfnamefont{A.}~\bibnamefont{Szameit}} \bibnamefont{and}
  \bibinfo{author}{\bibfnamefont{S.}~\bibnamefont{Nolte}}, \bibinfo{journal}{J.
  Phys. B: At. Mol. Phys.} \textbf{\bibinfo{volume}{43}},
  \bibinfo{pages}{163001} (\bibinfo{year}{2010}).

\bibitem[{\citenamefont{Ablowitz et~al.}(2015)\citenamefont{Ablowitz, Curtis,
  and Ma}}]{ACM15}
\bibinfo{author}{\bibfnamefont{M.~J.} \bibnamefont{Ablowitz}},
  \bibinfo{author}{\bibfnamefont{C.~W.} \bibnamefont{Curtis}},
  \bibnamefont{and} \bibinfo{author}{\bibfnamefont{Y.-P.} \bibnamefont{Ma}},
  \bibinfo{journal}{2D Mater.} \textbf{\bibinfo{volume}{2}},
  \bibinfo{pages}{024003} (\bibinfo{year}{2015}).

\bibitem[{\citenamefont{Rechtsman
  et~al.}(2013{\natexlab{b}})\citenamefont{Rechtsman, Plotnik, Zeuner, Song,
  Chen, Szameit, and Segev}}]{RPZ13}
\bibinfo{author}{\bibfnamefont{M.}~\bibnamefont{Rechtsman}},
  \bibinfo{author}{\bibfnamefont{Y.}~\bibnamefont{Plotnik}},
  \bibinfo{author}{\bibfnamefont{J.}~\bibnamefont{Zeuner}},
  \bibinfo{author}{\bibfnamefont{D.}~\bibnamefont{Song}},
  \bibinfo{author}{\bibfnamefont{Z.}~\bibnamefont{Chen}},
  \bibinfo{author}{\bibfnamefont{A.}~\bibnamefont{Szameit}}, \bibnamefont{and}
  \bibinfo{author}{\bibfnamefont{M.}~\bibnamefont{Segev}},
  \bibinfo{journal}{Phys. Rev. Lett.} \textbf{\bibinfo{volume}{111}},
  \bibinfo{pages}{103901} (\bibinfo{year}{2013}{\natexlab{b}}).

\bibitem[{\citenamefont{Rechtsman
  et~al.}(2013{\natexlab{c}})\citenamefont{Rechtsman, Zeuner, Plotnik, Lumer,
  Podolsky, Dreisow, Nolte, Segev, and Szameit}}]{RZP13}
\bibinfo{author}{\bibfnamefont{M.}~\bibnamefont{Rechtsman}},
  \bibinfo{author}{\bibfnamefont{J.}~\bibnamefont{Zeuner}},
  \bibinfo{author}{\bibfnamefont{Y.}~\bibnamefont{Plotnik}},
  \bibinfo{author}{\bibfnamefont{Y.}~\bibnamefont{Lumer}},
  \bibinfo{author}{\bibfnamefont{D.}~\bibnamefont{Podolsky}},
  \bibinfo{author}{\bibfnamefont{F.}~\bibnamefont{Dreisow}},
  \bibinfo{author}{\bibfnamefont{S.}~\bibnamefont{Nolte}},
  \bibinfo{author}{\bibfnamefont{M.}~\bibnamefont{Segev}}, \bibnamefont{and}
  \bibinfo{author}{\bibfnamefont{A.}~\bibnamefont{Szameit}},
  \bibinfo{journal}{Nature} \textbf{\bibinfo{volume}{496}},
  \bibinfo{pages}{196} (\bibinfo{year}{2013}{\natexlab{c}}).

\bibitem[{\citenamefont{Ozawa and Carusotto}(2014)}]{OC14}
\bibinfo{author}{\bibfnamefont{T.}~\bibnamefont{Ozawa}} \bibnamefont{and}
  \bibinfo{author}{\bibfnamefont{I.}~\bibnamefont{Carusotto}},
  \bibinfo{journal}{Phys. Rev. Lett.} \textbf{\bibinfo{volume}{112}},
  \bibinfo{pages}{133902} (\bibinfo{year}{2014}).

\bibitem[{\citenamefont{Berry}(1984)}]{Be84}
\bibinfo{author}{\bibfnamefont{M.}~\bibnamefont{Berry}},
  \bibinfo{journal}{Proc. Roy. Soc. London, Ser. A}
  \textbf{\bibinfo{volume}{392}}, \bibinfo{pages}{45} (\bibinfo{year}{1984}).

\bibitem[{\citenamefont{Xiao et~al.}(2010)\citenamefont{Xiao, Chang, and
  Nui}}]{XCN10}
\bibinfo{author}{\bibfnamefont{D.}~\bibnamefont{Xiao}},
  \bibinfo{author}{\bibfnamefont{M.-C.} \bibnamefont{Chang}}, \bibnamefont{and}
  \bibinfo{author}{\bibfnamefont{N.}~\bibnamefont{Nui}}, \bibinfo{journal}{Rev.
  Mod. Phys.} \textbf{\bibinfo{volume}{82}}, \bibinfo{pages}{1959}
  (\bibinfo{year}{2010}).

\bibitem[{\citenamefont{Cominotti and Carusotto}(2013)}]{CC13}
\bibinfo{author}{\bibfnamefont{M.}~\bibnamefont{Cominotti}} \bibnamefont{and}
  \bibinfo{author}{\bibfnamefont{I.}~\bibnamefont{Carusotto}},
  \bibinfo{journal}{Europhys. Lett.} \textbf{\bibinfo{volume}{103}},
  \bibinfo{pages}{10001} (\bibinfo{year}{2013}).

\bibitem[{\citenamefont{Kobayashi et~al.}(2011)\citenamefont{Kobayashi,
  Suzumura, Pi\'echon, and Montambaux}}]{KSPM11}
\bibinfo{author}{\bibfnamefont{A.}~\bibnamefont{Kobayashi}},
  \bibinfo{author}{\bibfnamefont{Y.}~\bibnamefont{Suzumura}},
  \bibinfo{author}{\bibfnamefont{F.}~\bibnamefont{Pi\'echon}},
  \bibnamefont{and}
  \bibinfo{author}{\bibfnamefont{G.}~\bibnamefont{Montambaux}},
  \bibinfo{journal}{Phys. Rev. B} \textbf{\bibinfo{volume}{84}},
  \bibinfo{pages}{075450} (\bibinfo{year}{2011}).

\bibitem[{\citenamefont{Tarruell et~al.}(2012)\citenamefont{Tarruell, Greif,
  Jotzu, Uehlinger, and Esslinger}}]{TGUJE12}
\bibinfo{author}{\bibfnamefont{L.}~\bibnamefont{Tarruell}},
  \bibinfo{author}{\bibfnamefont{D.}~\bibnamefont{Greif}},
  \bibinfo{author}{\bibfnamefont{G.}~\bibnamefont{Jotzu}},
  \bibinfo{author}{\bibfnamefont{T.}~\bibnamefont{Uehlinger}},
  \bibnamefont{and}
  \bibinfo{author}{\bibfnamefont{T.}~\bibnamefont{Esslinger}},
  \bibinfo{journal}{Nature} \textbf{\bibinfo{volume}{483}},
  \bibinfo{pages}{302} (\bibinfo{year}{2012}).

\bibitem[{\citenamefont{Jotzu et~al.}(2014)\citenamefont{Jotzu, Messer,
  Desbuquois, Lebrat, Uehlinger, Greif, and Esslinger}}]{JMDLUGE14}
\bibinfo{author}{\bibfnamefont{G.}~\bibnamefont{Jotzu}},
  \bibinfo{author}{\bibfnamefont{M.}~\bibnamefont{Messer}},
  \bibinfo{author}{\bibfnamefont{R.}~\bibnamefont{Desbuquois}},
  \bibinfo{author}{\bibfnamefont{M.}~\bibnamefont{Lebrat}},
  \bibinfo{author}{\bibfnamefont{T.}~\bibnamefont{Uehlinger}},
  \bibinfo{author}{\bibfnamefont{D.}~\bibnamefont{Greif}}, \bibnamefont{and}
  \bibinfo{author}{\bibfnamefont{T.}~\bibnamefont{Esslinger}},
  \bibinfo{journal}{Nature} \textbf{\bibinfo{volume}{515}},
  \bibinfo{pages}{237} (\bibinfo{year}{2014}).

\bibitem[{\citenamefont{Aidelsburger et~al.}(2015)\citenamefont{Aidelsburger,
  Lohse, Schweizer, Atala, Barreiro, Nascimb\`{e}ne, Cooper, Bloch, and
  Goldman}}]{ALSABNCBG15}
\bibinfo{author}{\bibfnamefont{M.}~\bibnamefont{Aidelsburger}},
  \bibinfo{author}{\bibfnamefont{M.}~\bibnamefont{Lohse}},
  \bibinfo{author}{\bibfnamefont{C.}~\bibnamefont{Schweizer}},
  \bibinfo{author}{\bibfnamefont{M.}~\bibnamefont{Atala}},
  \bibinfo{author}{\bibfnamefont{J.~T.} \bibnamefont{Barreiro}},
  \bibinfo{author}{\bibfnamefont{S.}~\bibnamefont{Nascimb\`{e}ne}},
  \bibinfo{author}{\bibfnamefont{N.~R.} \bibnamefont{Cooper}},
  \bibinfo{author}{\bibfnamefont{I.}~\bibnamefont{Bloch}}, \bibnamefont{and}
  \bibinfo{author}{\bibfnamefont{N.}~\bibnamefont{Goldman}},
  \bibinfo{journal}{Nat. Phys.} \textbf{\bibinfo{volume}{11}},
  \bibinfo{pages}{162} (\bibinfo{year}{2015}).

\bibitem[{\citenamefont{Duca et~al.}(2015)\citenamefont{Duca, Li, Reitter,
  Bloch, Schleier-Schmith, and Schneider}}]{DLRBSS15}
\bibinfo{author}{\bibfnamefont{L.}~\bibnamefont{Duca}},
  \bibinfo{author}{\bibfnamefont{T.}~\bibnamefont{Li}},
  \bibinfo{author}{\bibfnamefont{M.}~\bibnamefont{Reitter}},
  \bibinfo{author}{\bibfnamefont{I.}~\bibnamefont{Bloch}},
  \bibinfo{author}{\bibfnamefont{M.}~\bibnamefont{Schleier-Schmith}},
  \bibnamefont{and}
  \bibinfo{author}{\bibfnamefont{U.}~\bibnamefont{Schneider}},
  \bibinfo{journal}{Science} \textbf{\bibinfo{volume}{347}},
  \bibinfo{pages}{288} (\bibinfo{year}{2015}).

\bibitem[{\citenamefont{Fl\"aschner et~al.}(2015)\citenamefont{Fl\"aschner,
  Rem, Tarnowski, Vogel, L\"uhmann, Sengstock, and Weitenberg}}]{FRTVLSW15}
\bibinfo{author}{\bibfnamefont{N.}~\bibnamefont{Fl\"aschner}},
  \bibinfo{author}{\bibfnamefont{B.~S.} \bibnamefont{Rem}},
  \bibinfo{author}{\bibfnamefont{M.}~\bibnamefont{Tarnowski}},
  \bibinfo{author}{\bibfnamefont{D.}~\bibnamefont{Vogel}},
  \bibinfo{author}{\bibfnamefont{D.-S.} \bibnamefont{L\"uhmann}},
  \bibinfo{author}{\bibfnamefont{K.}~\bibnamefont{Sengstock}},
  \bibnamefont{and}
  \bibinfo{author}{\bibfnamefont{C.}~\bibnamefont{Weitenberg}}
  (\bibinfo{year}{2015}), \urlprefix\url{arXiv:1509.05763}.

\bibitem[{\citenamefont{Chang and Niu}(1996)}]{CN96}
\bibinfo{author}{\bibfnamefont{M.-C.} \bibnamefont{Chang}} \bibnamefont{and}
  \bibinfo{author}{\bibfnamefont{Q.}~\bibnamefont{Niu}},
  \bibinfo{journal}{Phys. Rev. B} \textbf{\bibinfo{volume}{53}},
  \bibinfo{pages}{7010} (\bibinfo{year}{1996}).

\bibitem[{\citenamefont{Shima and Nakayama}(1999)}]{SN99}
\bibinfo{author}{\bibfnamefont{H.}~\bibnamefont{Shima}} \bibnamefont{and}
  \bibinfo{author}{\bibfnamefont{T.}~\bibnamefont{Nakayama}},
  \bibinfo{journal}{Phys. Rev. B} \textbf{\bibinfo{volume}{60}},
  \bibinfo{pages}{14066} (\bibinfo{year}{1999}).

\bibitem[{\citenamefont{Diener et~al.}(2003)\citenamefont{Diener, Dudarev,
  Sundaram, and Niu}}]{DSND03}
\bibinfo{author}{\bibfnamefont{R.~B.} \bibnamefont{Diener}},
  \bibinfo{author}{\bibfnamefont{A.~M.} \bibnamefont{Dudarev}},
  \bibinfo{author}{\bibfnamefont{G.}~\bibnamefont{Sundaram}}, \bibnamefont{and}
  \bibinfo{author}{\bibfnamefont{Q.}~\bibnamefont{Niu}} (\bibinfo{year}{2003}),
  \bibinfo{note}{arXiv:cond-mat/0306184}.

\bibitem[{\citenamefont{Price and Cooper}(2012)}]{PC12}
\bibinfo{author}{\bibfnamefont{H.~M.} \bibnamefont{Price}} \bibnamefont{and}
  \bibinfo{author}{\bibfnamefont{N.~R.} \bibnamefont{Cooper}},
  \bibinfo{journal}{Phys. Rev. A} \textbf{\bibinfo{volume}{85}},
  \bibinfo{pages}{033620} (\bibinfo{year}{2012}).

\bibitem[{\citenamefont{Price and Cooper}(2013)}]{PC13}
\bibinfo{author}{\bibfnamefont{H.~M.} \bibnamefont{Price}} \bibnamefont{and}
  \bibinfo{author}{\bibfnamefont{N.~R.} \bibnamefont{Cooper}},
  \bibinfo{journal}{Phys. Rev. Lett.} \textbf{\bibinfo{volume}{111}},
  \bibinfo{pages}{220407} (\bibinfo{year}{2013}).

\bibitem[{\citenamefont{Sundaram and Niu}(1999)}]{SN99b}
\bibinfo{author}{\bibfnamefont{S.}~\bibnamefont{Sundaram}} \bibnamefont{and}
  \bibinfo{author}{\bibfnamefont{Q.}~\bibnamefont{Niu}},
  \bibinfo{journal}{Phys. Rev. B} \textbf{\bibinfo{volume}{59}},
  \bibinfo{pages}{14915} (\bibinfo{year}{1999}).

\bibitem[{\citenamefont{Fuchs et~al.}(2010)\citenamefont{Fuchs, Piechon,
  Goerbig, and Montambaux}}]{FPG10}
\bibinfo{author}{\bibfnamefont{J.-N.} \bibnamefont{Fuchs}},
  \bibinfo{author}{\bibfnamefont{F.}~\bibnamefont{Piechon}},
  \bibinfo{author}{\bibfnamefont{M.~O.} \bibnamefont{Goerbig}},
  \bibnamefont{and}
  \bibinfo{author}{\bibfnamefont{G.}~\bibnamefont{Montambaux}},
  \bibinfo{journal}{Eur. Phys. J. B} \textbf{\bibinfo{volume}{77}},
  \bibinfo{pages}{351} (\bibinfo{year}{2010}).

\bibitem[{\citenamefont{Castro~Neto et~al.}(2009)\citenamefont{Castro~Neto,
  Guinea, Peres, Novoselov, and Geim}}]{CGPNG09}
\bibinfo{author}{\bibfnamefont{A.~H.} \bibnamefont{Castro~Neto}},
  \bibinfo{author}{\bibfnamefont{F.}~\bibnamefont{Guinea}},
  \bibinfo{author}{\bibfnamefont{N.~M.~R.} \bibnamefont{Peres}},
  \bibinfo{author}{\bibfnamefont{K.~S.} \bibnamefont{Novoselov}},
  \bibnamefont{and} \bibinfo{author}{\bibfnamefont{A.~K.} \bibnamefont{Geim}},
  \bibinfo{journal}{Rev. Mod. Phys.} \textbf{\bibinfo{volume}{81}},
  \bibinfo{pages}{109} (\bibinfo{year}{2009}).

\bibitem[{\citenamefont{Montambaux
  et~al.}(2009{\natexlab{a}})\citenamefont{Montambaux, Pi\'echon, Fuchs, and
  Goerbig}}]{MPFG09a}
\bibinfo{author}{\bibfnamefont{G.}~\bibnamefont{Montambaux}},
  \bibinfo{author}{\bibfnamefont{F.}~\bibnamefont{Pi\'echon}},
  \bibinfo{author}{\bibfnamefont{J.-N.} \bibnamefont{Fuchs}}, \bibnamefont{and}
  \bibinfo{author}{\bibfnamefont{M.~O.} \bibnamefont{Goerbig}},
  \bibinfo{journal}{Eur. Phys. J. B} \textbf{\bibinfo{volume}{72}},
  \bibinfo{pages}{509} (\bibinfo{year}{2009}{\natexlab{a}}).

\bibitem[{\citenamefont{Montambaux
  et~al.}(2009{\natexlab{b}})\citenamefont{Montambaux, Pi\'echon, Fuchs, and
  Goerbig}}]{MPFG09b}
\bibinfo{author}{\bibfnamefont{G.}~\bibnamefont{Montambaux}},
  \bibinfo{author}{\bibfnamefont{F.}~\bibnamefont{Pi\'echon}},
  \bibinfo{author}{\bibfnamefont{J.-N.} \bibnamefont{Fuchs}}, \bibnamefont{and}
  \bibinfo{author}{\bibfnamefont{M.~O.} \bibnamefont{Goerbig}},
  \bibinfo{journal}{Phys. Rev. B} \textbf{\bibinfo{volume}{80}},
  \bibinfo{pages}{153412} (\bibinfo{year}{2009}{\natexlab{b}}).

\bibitem[{\citenamefont{Bellec et~al.}(2013)\citenamefont{Bellec, Kuhl,
  Montambaux, and Mortessagne}}]{BKMM13}
\bibinfo{author}{\bibfnamefont{M.}~\bibnamefont{Bellec}},
  \bibinfo{author}{\bibfnamefont{U.}~\bibnamefont{Kuhl}},
  \bibinfo{author}{\bibfnamefont{G.}~\bibnamefont{Montambaux}},
  \bibnamefont{and}
  \bibinfo{author}{\bibfnamefont{F.}~\bibnamefont{Mortessagne}},
  \bibinfo{journal}{Phys. Rev. Lett.} \textbf{\bibinfo{volume}{110}},
  \bibinfo{pages}{033902} (\bibinfo{year}{2013}).

\end{thebibliography}
\end{document}